# Evolutionary Ecology Models of Weed Life History

Jack Dekker

Weed Biology Laboratory, Agronomy Department, Iowa State University, Ames, Iowa 50011
Email:  jdekker@iastate.edu

*Keywords*:  *Setaria*; seedling recruitment; seedling emergence; natural selection; weedy adaptation; informational content; model inference; model prediction; demographic models; population dynamics models; opportunity spacetime

TABLE OF CONTENTS:

**ABSTRACT**.  Weeds and invasive plants perform the colonization niche by seizing locally available opportunity spacetime created by human activity.  The urge to understand and predict weed life history behavior provides a strong scientific and practical motivation to develop models.  Most current weed models are quantitative and demographic.  This chapter is a critical review of the limitations of demographic models and well as the opportunities provided by evolutionary models.  Several fundamental flaws are associated with the way the local population is represented in demographic models.  The first artifact is the confounding effects of plant, as opposed to animal, population structure.  The second derives from how unique individual phenotypes in the local population are represented.  The third arises from population membership changes with evolutionary time that compromise assumptions of deme covariance structure.  As an alternative to demographic models, an evolutionary model of weed population dynamics is based on the actions of functional traits guided weedy plant life history behavior in a deme as a consequence of natural selection and reproductive success among excess variable phenotypes in response to the structure, quality and timing of locally available opportunity spacetime.  The thesis of this chapter is that understanding population dynamics in agroecosystems requires a qualitative evolutionary representation of local populations based upon the two component processes of natural selection and elimination resulting in weedy adaptation. FoxPatch is an example of an evolutionary model based on the processes of natural selection: generation of variation, selection and elimination. FoxPatch represents weedy *Setaria* seed-seedling life history population dynamics with explicit seed process prediction rules via trait-process-signal modeling. Inherent, trait-based processes are modulated by effective signals ($O_2$-$H_2O$ -thermal-time) determining soil seed behavior. Phenotypic variation is generated during embryogenesis by induction of variable seed germinability-dormancy capacities among parental offspring, seed heteroblasty. Seed heteroblasty, modulated by $O_2$-$H_2O$-thermal-time, thereafter determines reversible seasonal dormancy cycling in the soil as well as irreversible germination leading to seedling emergence. Hedge-bet patterns of seedling emergence exploit predictable local opportunity spacetime (resources, conditions, cropping disturbances, neighbors). There exists a relationship between seed heteroblasty at abscission and its subsequent behavior in the soil that can be exploited to predict recruitment pattern: seed heteroblasty 'blueprints' seedling emergence pattern.

## INTRODUCTION

"The existence of two levels of population structure in plants makes for difficulties, but the problems are much greater if their existence is ignored. One of the strongest reasons why a population biology of plants failed to develop alongside that of animals was that counting plants gives so much less information than counting animals. A count of the number of rabibits or *Drosophila* or voles or flour beetles gives a lot of information: it permits *rough* predictions of population growth rates, biomass and even productivity. A count of the number of plants in an area gives extraordinarily little information unless we are also told their size. Individual plants are so "plastic" that variations of 50,000-fold in weight or reproductive capacity are easily found in individuals of the same age. Clearly, counting plants is not enough to give a basis for a useful demography. The plasticity of plants lies, howerever, almost entirely in variations of the number of their parts. The other closely related reason why plant demography has been slow to develop is that the clonal spread of plants and the break-up of old clones often makes it impossible to count the number of genetic individuals.

The problems are great: they can be regarded as insoluable and a demography of plants an unattainable ideal, or they can be ignored with a certainty of serious misinterpretation, or they can be grasped and methods, albeit crude, developed to handle the problem." (Harper, 1977)

Weedy and invasive plants perform the plant colonization niche. Weedy plants are the first to seize and exploit the opportunity spacetime created by human disturbance, notably in resource-rich agricultural cropping systems. The urge to understand and predict weed life history behavior with time has provided a strong scientific and practical motivation for the development of these models. Weed models are tools with the potential to provide improved scientific understanding of changing weed populations, including insights into the biological functioning of these plants, and prediction of future life history population dynamics. Weed modeling can also provide practical support for crop management decision making, including evaluation of weed management tactics and strategy, risk, economics and efficacy. Modeling can also be a less expensive means of providing information compared to that of field experimentation. Much progress has been made to realize the potential of weed modeling, but much remains undone.

The basis of most current weed models is quantitative and demographic: comparisons over years of the numbers and sizes of plants per unit area at different times of their life history. The current state of affairs, including limitations of current demographic models, have been featured in two recent reviews in Weed Research (Holst et al., 2007; Freckleton and Stephens, 2009).

The opportunities and limitations of models arise from the manner in which weeds and their life histories are represented, the inferences that can be derived from the informational content of the models, and the consequences of these factors on the ability of the model to predict future behavior. The purpose of this review is to assess the limits and potentials of two different, but compatible, types of weed population dynamics models: demographic, and those based on functional phenotypic traits and the biological processes of natural selection, elimination and evolutionary adaptation. Models of both types are assessed in terms of how they represent weed life histories as well as ability to infer and predict future behavior based on their inherent informational content.

**Weed Life History Models**

A model is a representation of reality. It is inherently an abstraction and a simplification. It is a conceptual framework of a system constructed by indicating which elements should be represented and how these elements are interrelated. This conceptual framework then is translated into algorithms, precisely defined step-by-step procedures by which dynamics are carried out. What elements should be represented in a weed population dynamics model? The first, most important element is a group of plants of one weed species occupying a local habitat. This population is usually isolated to some degree from other populations, but local populations over spatial scales of landscape to global interact (e.g. gene flow) with each other to form metapopulations. It is the local population, the deme, that is the unit of evolution. Populations change with time. Population dynamics are changes in the quality and quantity of member phenotypes, as well as the biological and environmental processes influencing those changes.

What conceptual framework can best represent the interrelationships among members of a population? One crucial component of any conceptual framework of weed models is the life history of the weed species: "The life cycle is the fundamental unit of description of the organism." (Caswell, 2001). Holst et al. (2007) also conclude that "Almost all models consist of a number of life cycle stages, nearly always including at least seeds, seedlings and mature plants." (figure 1).

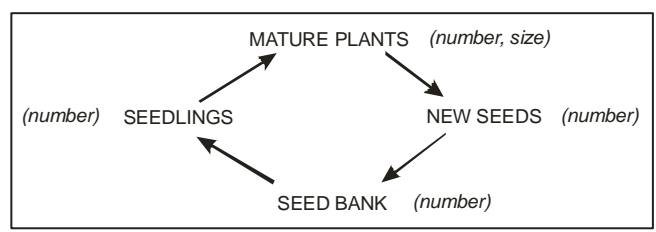

**Figure 1.** Representation of the an annual weed species life history by discrete life history phases, or states (mature plant, new seeds, seed bank, seedlings) and the metrics used for their measurement (number, size); arrows represent transitions between states; redrawn from Holst et al., 2007.

A weed population dynamics model is a representation of the phenomena of a weed population's life history growth and development, from fertilization to death. The model in figure 1 represents weed life history phases as discrete phenotypic states of the individual organism with growth. The demographic form of a life history model is quantifiable, with measurements of changes in phase state pool number and sizes with time, often expressed on a unit area basis. What this model does not contain are the deterministic biological processes that drive growth and development during life history. These uncharacterized processes are represented as transitions between quantitative state pools (figure 1, arrows).

**Demographic Weed Life History Population Dynamics Models**

Representations of weed life history population dynamics have largely been accomplished using demographic models (see Freckleton and Stevens, 2009; Holst, et al., 2007):

> "The essence of population biology is captured by a simple equation that relates the numbers per unit area of an organism $N_t$ at some time $t$ to the numbers $N_{t-1}$ one year earlier." (Silvertown & Doust, 1993)

From this perspective, weed population dynamics can be represented in its most essential form by this function describing the interrelationships of elements:

$$N_t = N_{t-1} + B - D + I - E$$

Where: $N_t$, number per unit area organisms at time $t$; $N_{t-1}$, number per unit area organisms one year later; B, number of births; D, number of deaths; I, immigrants in; E, emigrants out. Schematically translating this demographic function onto figure 1 reveals four potential life history state phases, or pools, and the relationship of the life history to its complementary metapopulation (I, E), as well as to mortality (D; figure 2).

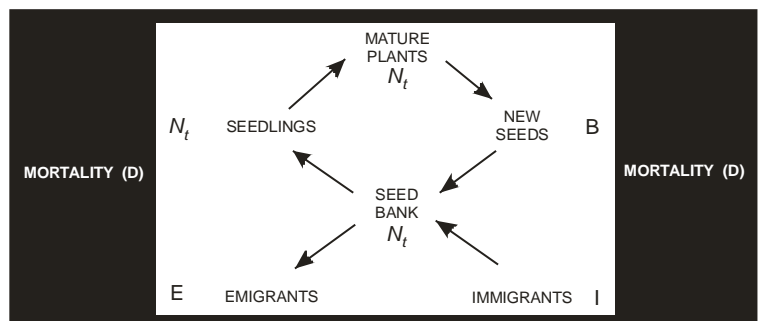

**Figure 2.** Demographic representation of an annual weed species life history by quantification (number, size) of discrete life history states (mature plant, new seeds, seed bank, seedlings); population size influenced by metapopulaton immigration and emigration dispersal events into and out of the soil seed pool; arrows represent transitions between states.

Weed population dynamics are algorithmically represented by the calculation of lambda ($\lambda$), the rate of population size change over one generation where: $\lambda = R_0$, net reproductive rate, rate of population increase over a generation; $\lambda_t = N_t / N_{t-1}$, annual population growth rate, finite rate of increase. The finite rate of increase for a population is also expressed as a measure of $W$, so-called Darwinian fitness. The most common formulation of weed population models is as an iterative equation with next year's population calculated from that of the current year (Holst, et al., 2007). The rate of population change over several generations is schematically represented in figure 3.

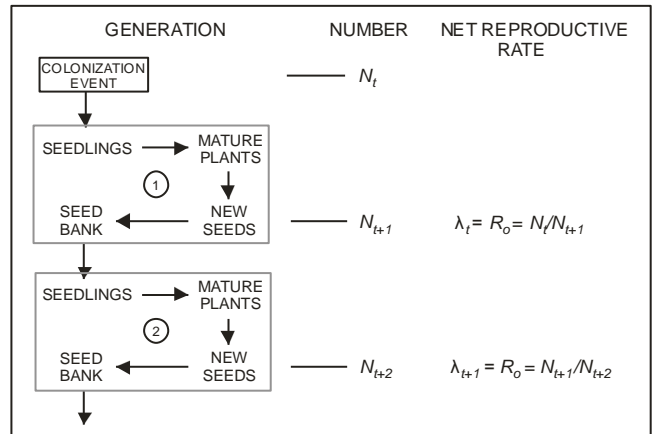

**Figure 3** Change in numbers or organisms per unit area ($N_t$) with life history generation time ($N_t$, $N_{t-1}$, etc.) and associated net reproductive rates ($\lambda = R_0$) per generation (1, 2, 3, 4).

The simplest form of demographic models presented above has been refered to as the "standard population model" (Holst et al., 2007; Sagar and Mortimer, 1976). This simple form has been expanded and extended in several ways. Variation on this basic theme include genetic or spatial components or aspects. The primary amplification of these demographic models occurs in increased attention to life history states: seedling recruitment, mature plants, new seeds and the soil seed bank. Interactions among plants is typically expressed in terms of final biomass of crop and weeds at harvest, with various shortcuts utilized predict the outcome in terms of yield. In most models the outcome of this process is simply reflected in a single competition parameter, which is kept constant over the years. Stochasticity is introduced in some of the models to generate a more irregular population development. Quantification of new seed is based on fecundity per unit area, and is derived in various ways (Holst et al., 2007). Little is said in Holst et al. (2007) about soil seed pools, which appear to be equated with quantities of old and new seeds, immigrant and emigrant seeds, all with similar or uniform qualities.

Given that the current state of the art of weed population dynamics modeling is represented almost entirely with demographic models, what inherent properties of quantitative models provide the ability to predict future weed behavior? What alternatives exist to overcome the limitations of quantitative models?

## DISCUSSION

**Representation and Information, Inference and Prediction**
The opportunities and limitations of models arise from the manner in which weeds and their life histories are represented, the inferences that can be derived from the informational content of the models, and the consequences of these factors on the ability of the model to predict future behavior.
**Representation and information**. Information is the meaning given to data (facts, norms) by the way in which it is interpreted. It is a message received and understood.

"Information is revealed in the correlation between two things that is produced by a lawful process (as opposed to coming about by sheer chance). Information itself is nothing special. It is found wherever causes leave effects. What is special is information processing. We can regard a piece of matter that carries information about some state of affairs as a symbol: it can "stand for" that state of affairs. As a piece of matter, it can do other things as well, physical things, whatever that kind of matter in that kind of state can do according to the laws of physics and chemistry." (Pinker, 1997).

Information can be viewed as a type of input important to the function of the organism (e.g. resources), causal inputs. Information is captured in the physical structure of the organism that symbolize things in nature and that respond to external stimuli allowing causation. Information exists in the physical and physiological structures that capture the way a complex organism can tune itself to unpredictable aspects of the world and take in the kinds of stimuli it needs to function (Pinker, 1997). A good example of biological information taking a physical form is the genome, the informational content of DNA. Closer to home for weed modelers, it is contained in the functional traits of a weed phenotype. Functional traits, such as seed dormancy capacity (e.g. see discussion below on seed heteroblasty), respond to the environment in a particular manner during life history to maximize survival and reproduction. Information can take other forms too.

For weed modelers, what is information? How is information represented in a weed model? Different models contain differing amounts of useful information, the basis of inference and prediction. Inference and prediction are restricted to the informational content of the model.

In the demographic model presented above, the phenotypic identity of weed plants in a local population is represented in their numbers and sizes in spacetime. The informational content derives from the meaning and interpretation given to the numbers and sizes of plants of a particular weed species observed in a particular place, at a particular time (season, life history phase). These metrics provide no information about causation or dynamics. Causation can be inferred indirectly if the same plant in the same location is observed at a later time.

In the evolutionary model presented below, phenotypic identity and informational content are contained in the variation in germination-dormancy capacities induced by a parent plant during embryogenesis, seed heteroblasty. Seed heteroblasty is information physically captured in the structure of the various seeds. It is the behavioral blueprint that responds to specific environmental signals in the soil resulting in seedling emergence carefully timed to the historical occurrence of predictable cropping system disturbances (e.g. herbicide application, tillage, harvesting). Seed heteroblasty is the physical information encoded in the morpho-physiology of the seed, it is the cause and determinant of its subsequent life history. It acquired this preadapted physical information by the processes of natural selection-elimination over time. Causation for its life history can be directly inferred from a seed germination assay at seed abscission.

Evolutionary models represent weed life history dynamics in a local population by capturing the physical and behavior information contained in functional traits of the individual weed phenotype that respond to specific environmental signals, opportunity spacetime, in a manner that optimizes their fitness in terms of survival and reproduction.

A complementary way of measuring informational content of a weed model is provided by algorithmic information theory, which measures the information content of a list of symbols based on how predictable they are, or more specifically how easy it is to compute the list through a program. A symbol is an entity with two properties glued together. This symbol carries information, and it causes things to happen. When the caused things themselves carry information, we call the whole system an information processor, or a computer. The processing of symbols involves arrangements of matter that have both representational and causal properties, they simultaneously carry information about something and take part in a chain of physical events. Those events make up a computation (Pinker, 1997). Algorithmically, it is a set of rules that provides an accurate and complete description of a life history capturing its key properties. It provides an algorithmic, or computational, means of forecasting the future behavior of that life history.

Weed phenotypes contain these 'symbols' in the physical form of the functional traits they possess. For example, the DNA coding for the multiple traits of phenotypic plasticity in a weed species is information. It is a physical algorithm computed by the phenotype at every step of its life history that results in its current form and function closely tracking the environmental signals it receives in the local community. Model representations can contain algorithmic forms of this type of information: step-by-step recipes that will, when given an initial state, proceed through a well-defined series of successive states, eventually terminating in an end-state.

A truly dynamic weed population model then would be one that incorporates model algorithms that specify the life history steps a phenotype will go through given an initial state (e.g. seed heteroblasty; time of emergence), as modulated by the specific environment encountered by the individual phenotype in a local population. The biological foundation of these model algorithms is the manner in which specific functional traits are represented.

Evaluating a weed population dynamics model should include a search for its informational content, specifically its representation of the biological traits of the phenotypes of a local population that determine its life history trajectory to survive and reproduce.

**Inference**. Inference is the process of reasoning from premises to a conclusion, a deduction. The primary premise, or assumption, of demographic models is that the essence of population biology is captured by a simple equation that relates changes in numbers of organisms per unit area of space with time (Silvertown and Doust, 1993). With this premise, what inferences can, and cannot, be derived from a demographic model of weed population dynamics? A critical review of demographic models reveals a dearth of informational content, flaws in its representation of the deme and life history developmental behavior, and insufficient model formalization.

Several fundamental flaws are associated with the way the local population is represented in demographic models. The first artifact is the confounding effects of plant, as opposed to animal, population structure. The second derives from how unique individual phenotypes in the local population are represented. The third arises from population membership changes with time that compromise assumptions of deme covariance structure.

<u>Population structure</u>. The local population, the deme, is the fundamental unit of biological evolution. The deme consists of unique individual phenotypes, the units of natural selection. Of crucial importance is how these two components of any weed population dynamic model are represented. Plants respond in a highly plastic manner to locally available opportunity. Unlike animals, plant quantification fails to capture the qualities of the population that drive future dynamics. Demographic weed models that fail to represent this structural nature of plant populations are therefore compromised at conception. John Harper (1977, pp. 25-26) warned of this fatal flaw in weed models (see quote at beginning of this chapter). Demographic representation of the structure of a local plant population depends therefore on the specification of the number of individuals (level one), and on the number and variability of parts (e.g. leaves, tillers, meristems) of each individual (level two).

<u>Individual phenotypic identity</u>. Natural elimination acts by nonrandomly selecting the fittest individuals in the local population. Changes in demes therefore are an adaptive reflection of the unique biodiverse qualities of those survivors. Weed models are critically evaluated for their ability to represent individual phenotypic identity by means of their functional properties. Weeds assemble in local communities as collections of unique phenotypes. The urge to simplify their representation by categorical or average qualities obscures this biodiversity.

> "The assumptions of population thinking are diametrically opposed to those of the typologist. The populationist stresses the uniqueness of everything in the organic world. What is true for the human species – that no two individuals are alike – is equally true for all other species of animals and plants. Indeed, even the same individual changes continuously throughout its lifetime when placed in different environments. All organisms and organic phenomena are composed of unique features and can be described collectively only in statistical terms. Individuals, or any kind of organic entities, form populations of which we can determine the arithmetic mean and the statistics of variation. Averages are merely statistical abstractions, only the individuals of which the population are composed have reality. The ultimate conclusions of the population thinker and of the typologist are precisely the opposite. For the typologist, the type (*eidos*) is real and the variation an illusion, while for the populationist the type (average) is the abstraction and only the variation is real. No two ways of looking at nature could be more different."
> (Mayr, E. 1959)

Demographic representations of weed populations are limited to the extent that numbers of plants fail to provide information of the qualities of their members. The demographic representation of weed population dynamics is an incomplete abstraction because it ignores the importance of phenotypic variation by averaging behaviors at experimentally convenient times in life history. Measurement of quantities and sizes of uncharacterized phenotypes, and the uncharacterized processes of transitions between life history states, provide little inherent inference of population dynamics.

Local population dynamics. The third, and possibly the most telling, artifact of demographic representations of populations arises from the changing phenotypic structure of the local community with time. Natural selection eliminates lesser fit individuals to the enrichment of the survivors. As such the phenotypic-genotypic composition of the deme is constantly changing with time. During the growing season mortality alters the composition of the population. The population genetic structure of a local soil seed pool is different every year with the addition of offspring from those favored individuals. As such, demographic models represent populations as constant qualitative entities. Causation cannot be inferred from plant numbers that consist of different individual phenotypes. Inferences derived from them are incomplete as the assumptions underlying population covariance structure are violated. Covariance is a measure of how much two variables change together, for example plant number with time. The phenotypic membership of the local deme and soil seed pool changes as natural selection favors some and eliminates others. Natural selection violates this covariance structure by assuming the individuals are the same at each life history measurement time in the local habitat.

Life history development and behavior. Strong inferences of weed population dynamics can be made when the functional traits driving individual phenotypic behavior in local population are represented. Some of the most important functional traits of weeds are found in individual plant polymorphism and plasticity.

Life history states and processes. Demographic models represent weed populations by quantifying numbers of plants, their sizes and their density per unit area of space at discrete times in their life history. Lacking is a representation of the developmental growth processes that cause transitions between life history states to occur (arrows, figure 1). Demographic models do not embrace the dynamic processes causing weed life history, despite the claim that "Processes governing the transition from one stage to the other, like germination and seed production and processes responsible for the losses that occur throughout, like seed mortality, plant death and seed predation, are included." (Holst et al., 2007). Process is indirectly infered, a surrogate derived from computational number-size frequency transitions between discrete life history times. Mature plant number and size are not the competitive processes of interaction among neighbors. Soil seed pool numbers are not the motive forces driving the processes of germination, dormancy reinduction and seedling recruitment. New seed numbers do not reveal adaptive changes in these new phenotypes caused by natural selection and elimination: changes in the genetic-phenotypic composition of the local population. Life history developmental behavior is motivated by specific environmental signals stimulating functional traits inherent in the phenotype. Weed population dynamics come about as a direct adaptive consequence of generating phenotypic-trait variation among excess progeny in the deme, followed by the survival and reproduction of the fittest phenotypes among those offspring with time.

Polymorphism and plasticity. Individual weed phenotypes derive fitness from their heterogeneity by exploiting local opportunity. Weed population structure is difficult to model unless somatic polymorphism and phenotypic plasticity are represented, inherent functional traits that control life history behavior as well as allow the individual to assume a size and function appropriate to its local opportunity spacetime. Somatic polymorphism is the production of different plant parts, or different behaviors, within the individual that are expressed independently of its local environment. Seed heteroblasty is an example of parentally-induced dormancy heterogeneity among offspring that provides strong inferences of future behavior. Phenotypic plasticity is the capacity of a weedy plant to vary morphologically, physiologically or behaviorally as a result of environmental influences on the genotype during the plant's life history. Experimentally capturing this level of population structure entails measuring the population of phenotypes expressed by a single genotype when a trait changes continuously under different environmental and developmental conditions: the reaction norm. The reaction norm in population structure is expressed by number and size of constituent leaf, branch, flower and root modules of the individual plant that vary in response to the locally available opportunity spacetime. This plasticity of form confounds the ability of a purely demographic model to make predictions of population growth rates, biomass and even productivity. The consequence of phenotypic plasticity is that plants growing under density stress typically have a skewed distribution of individual plant weights, especially when they compete for light. Skewing of the frequency distribution (numbers of plants versus weight per plant) increases with time and with increasing density (plants per unit area). Typically at harvest a hierarchy of individuals is established: a few large dominants and a large number of suppressed, smaller, plants. The individual weeds in the hierarchical population structure possess the potential for explosive, nonlinear exponential growth and fecundity. Individual weed plants have the potential to produce a very large range of seed numbers depending on their size. The range in reproductive capacities of plants extends from 1 to $10^{10}$ (approaching infinity for vegetative clone propagule production; Harper, 1977). There exists a danger in assuming that the average plant performance represents the commonest type, or most typical, plant performance (Dekker, 2009).

Model formalization and measurement metrics.

*Hypotheses of local weed population dynamics*. Any model of weed behavior must be preceded by an experimental hypothesis of how population dynamics comes about: to what is the deme adapted? It should be a statement of an overarching intuition of how the biological system works, or the primary forces driving its expression. Such a hypothesis should appropriately begin with the intelligent designer of the system: human agricultural activity. Such a hypothesis could provide a tool to realistically guide the mathematical, algorithmic and statistical formalization of model components, metrics and output. No hypothesis of this type has been proposed for demographic models.

*Mathematical, algorithmic, statisitical model formalization and component description.* A model is a representation of reality. It is inherently an abstraction and a simplification. It is a conceptual framework of a system constructed by indicating which elements should be represented and how these elements are interrelated. This conceptual framework then is translated into algorithms, precisely defined step-by-step procedures by which dynamics are carried out. Many models are published without "… a complete description of the model logic and mathematics, including the parameter values."; of the 134 papers reviewed, 16-19% were not open for re-use or even critique. (Holst et al., 2007). Inference in simple and complex systems and models derives from the definition of model parameter space and algorithmic solutions of population dynamics:

> "An intelligent system, then, cannot be stuffed with trillions of facts. It must be equipped with a smaller list of core truths and a set of rules to deduce their implications." (Pinker, 1997)

> "The real issue here is the apparent reduction in simplicity. A skeptic worries about all the information necessary to specify all the unseen worlds. But an entire ensemble is often much simpler than one of its members. The principle can be stated more formally using the notion of algorithmic content." "… the whole set is actually simpler than a specific solution …" "The lesson is that complexity increases when we restrict our attention to one particular element in an ensemble, thereby losing the symmetry and simplicity that were inherent in the totality of all the elements taken together." (Tegmark, M. 2009)

> "Spatially explicit models tend to get complex, or mathematically demanding, like the model of neighbourhood interference between *Abutilon theophrasti* and *Amaranthus retroflexus* (Pacala & Silander, 1990). Another hindrance to fully grasp these models is that they may contain so many details, that it makes a full description of the model in scientific journals impossible, e.g. the within-field model of Richter *et al.* (2000) or the landscape model of Colbach *et al.* (2001b). To counteract this inherent complexity in spatial processes, one can reduce the complexity of the weed model itself. But this makes for very abstract models which can be difficult to relate to real weed population dynamics (e.g. Wang *et al.*, 2003)."
> "… the danger that the model develops into a monstrous specimen covering far too many facets and bearing an enormous parameter requirement. Collecting relevant parameters then becomes a time consuming exercise or might even develop into an objective on its own, putting the focus on analysis, rather than on synthesis of knowledge. Additionally, models containing too many parameters are often characterized by enormous error margins, and often lose their robustness." (Holst et al., 2007)

*Random-nonrandom processes*. Any model of weed population dynamics must accurately represent both random and nonrandom processes. Holst et al. (2007) indicate that stochastic models can be used to explain past population dynamics. If successful, stochastic models gain credibility as predictive tools of long-term population dynamics. The authors indicate that stochastic models are a tool to handle the uncertainty of future conditions. This review classifies environmental unpredictability, agricultural practice (cropping disturbances) and statistical error in model parameter estimates as random, unpredictable, and stochastic. Classification of some of these experimentally tractable phenomena (e.g. cropping disturbance; survival and reproduction) as random is inappropriate. Significantly for this review of evolutionary weed population dynamic models, they classify natural demographic variation in reproduction and mortality as stochastic. Apparently Charles Darwin's contributions (1859) are underappreciated by demographic weed modelers. Variational evolution of a population or species occurs through changes in its members by natural selection, the processes of nonrandom elimination and nonrandom sexual selection (Mayr, 2001).

**Predicting weed population dynamics**.

"It's hard to make predictions, especially about the future." (Yogi Berra).

Two recent reviews of weed modeling have come to similar conclusions (Holst et al., 2007; Freckleton and Stephens, 2009). Prediction is an emergent property of the inherent biological information contained within the individual weed phenotype (and its traits) as it accomplishes its life history survival and reproduction. Demographic models inherently do not contain this biological information. Limitations in the inferences that demographic models render them of limited utility in predicting future behavior.

The predictability of a model is based in its complexity. The work of nobel laureate F.A. Hayek (1974) is revealing. He distinguished the capacity to predict behavior in simple systems and those in complex systems through modeling. Complex biological phenomenon could not be modeled effectively in the same manner as those that dealt with essentially simple phenomena like physics. Complex phenomena, through modeling, can only allow pattern predictions, compared with precise predictions made of non-complex phenomena. How then is it possible to predict weed population dynamics? What is missing in demographic population models is the biological information contained in weedy traits whose expression drives the missing deterministic processes, processes incorrectly attempted to be replaced by stochastic probabilities of knowable weed phenomena (Holst et al., 2007).

What then are the "…smaller list of core truths and a set of rules to deduce their implications." (Pinker, 1997) that will simplify weed population models and allow strong inference and predictability? Intuitively, these core truths most come from the inherent biological traits of the weeds themselves. It is to this that evolutionary models are directed.

**Evolutionary, Trait-Based, Weed Life History Population Dynamics Models**

"As the famous geneticist T. Dobzhansky has said so rightly, "Nothing in biology makes sense, except in the light of evolution." Indeed, there is no other natural explanation than evolution for [biological phenomena]." (Mayr, 2001).

An evolutionary model of population changes based on the actions of functional traits might be guided by the following hypothesis. Weedy and invasive plants perform the plant colonization niche. Weedy plants are the first to seize and exploit the opportunity spacetime created by human disturbance, notably in resource-rich agricultural cropping systems. Local opportunity spacetime is the habitable space available to an organism at a particular time which includes its resources (e.g. light, water, nutrients, gases) and conditions (e.g. heat, climate, location), its disturbance history (e.g. tillage, herbicides, winter), and neighboring organisms (e.g. crops, other weed species). Therefore, it is hypothesized that weedy plant life history behavior in a deme is a consequence of natural selection and reproductive success among excess variable phenotypes (and functional traits) in response to the structure, quality and timing of locally available opportunity spacetime.

What alternative is there to quantitative demographic life history models to represent weed population dynamics? How can the limitations and artifacts of quantitative demographic models be overcome? How is the essence of population biology captured in a life history representation? The thesis of this chapter is that understanding population dynamics in agroecosystems requires a qualitative evolutionary representation of local populations based upon the two component processes of natural selection and elimination resulting in weedy adaptation. Evolutionary models based on the two component processes of natural selection (generation of variation, selection and elimination) are discussed in terms of these same critical factors.

**Weed population dynamics: consequences of the process of natural selection-elimination.** The essence of population biology is captured by a weed life history representation stated in the form of the processes of natural selection: the fittest parents generate phenotypic variation in their offspring that preferentially survive and reproduce in the local deme. Figure 1 can be redrawn to represent this in a much simplified form:

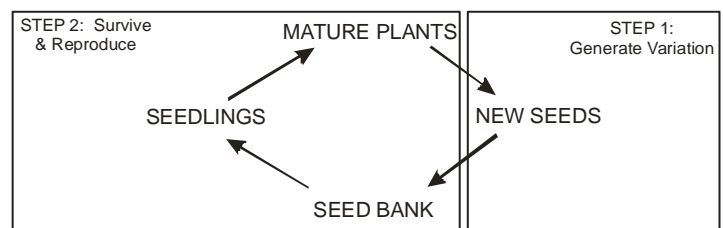

**Figure 4**. Representation of an annual weed species life history in terms of the two component processes of natural selection and elimination: step 1, production of phenotypic variation by the fittest parent plants; step 2, survival and reproduction of the fittest phenotypes, elimination of the others.

In each generation, new seed dispersed into the local deme comes from the fittest parent plants of the previous generation. In this view the phenotypic composition of the local deme is constantly changing during life history. Life history is not a repetitive cycle, but a spiral of overlapping life histories of changing individuals better adapted to the local habitat (figure 5).

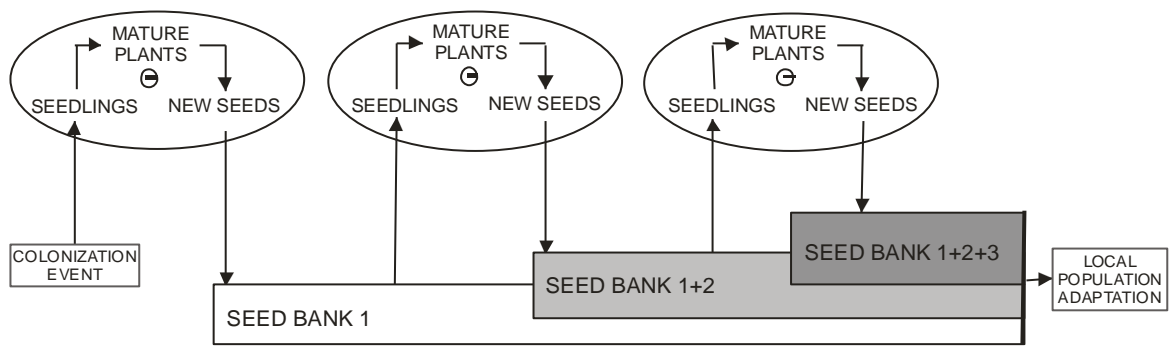

**Figure 5** Schematic representation of the adaptive changes in the local population of an annual weed species through several generations (life cycles) as a consequence of the two processes of natural selection and elimination.

Life history does not begin at the same starting condition with each new generation. New surviving seed join the preexisting seed pool in the soil to form the new local population every winter in a sexually reproducing annual weed species. The local population is dynamic, its phenotypic composition (the plant communities of the future) changes with new addition and loss. It is an expanding spiral for growing populations, a constricting spiral for dying populations. Therefore both the quantity and quality (traits) of the individual phenotypes in the deme change with time: adaptation to the local habitat. This evolutionary adaptation is the most dynamic element of the weed population that any realistic model must represent.

The phenotypic composition of each new local population changes with the recruitment of seedling from the seed pool. The composition of the seed pool is the dynamic element of local adaptation: the progeny of the fittest individuals selected from the previous generations. The representation of this changing seed pool is most challenging element in the formalization of a realistic life history model.

Sexually reproducing, annual, weed population dynamics are the adaptive consequence of natural selection and elimination of excess individuals in the local deme. This evolutionary process is represented by two processes and 5 conditions (Table 1)

| Precondition 1: Excess local phenotypes compete for limited opportunity spacetime | |
|---|---|
| Process step 1: Produce phenotypic variation | Condition 1: variation in individual traits |
| | Condition 2: variation in individual fitness |
| Process step 2: Survival and reproduction of the fittest phenotypes | Condition 3: survive to reproduce the fittest offspring, eliminate the others |
| | Condition 4: reproductive transmission parental traits to offspring |
| Adaptation arises in the local population of phenotypes | |

**Table 1**. The local adaption of a sexually reproducing weed population by the processes (and component conditions) of natural selection of the fittest phenotypes.

**The local habitat and opportunity spacetime**. Plants will fill any available and habitable growing space, therefore the primary resource limiting plant growth is habitable space. Every potentially habitable space includes the resources (e.g. relative abundance of light, water, nutrients, gases) and conditions (e.g. relative abundance of heat) of that location, its disturbance history, as well as the neighboring organisms that occupy that space. The structure of available and habitable space to an invading plant is also opportunity space at a particular time, opportunity spacetime.

FoxPatch, an evolutionary trait-based model of weed *Setaria* species-group life history is reviewed as an example of an alternate mode of representation providing the predictive ability of seed heterblasty blueprinting the crucial life history threshold events of seedling emergence.

**FoxPatch: A Trait-based, Natural Selection Process Representation of the *Setaria* Species-Group Seed-Seedling Life History Dynamics**

FoxPatch, an evolutionary trait-based model of weed *Setaria* species-group life history, is reviewed as an example of an alternate mode of representation utilizing the information contained in the seed heterblasty 'blueprint' to predict the crucial life history threshold events of seedling emergence. FoxPatch represents life history dynamics with two nested process models. Overarching natural selection processes are defined by the functional traits responsible for the processes of seed-seedling life history development (table 1).
FoxPatch represents weedy *Setaria* species-group (*S. viridis*, *S. verticillata*, *S. pumila*, *S. geniculata*; Dekker, 2003) life history, but is experimentally focused on *S. faberi*. Annual, self-fertilizing *Setaria* weed life history is represented with five life history states (1-5) and six developmental processes (A-F) (figure 6) (Dekker et al., 2003). The risk of death is constant during life history.

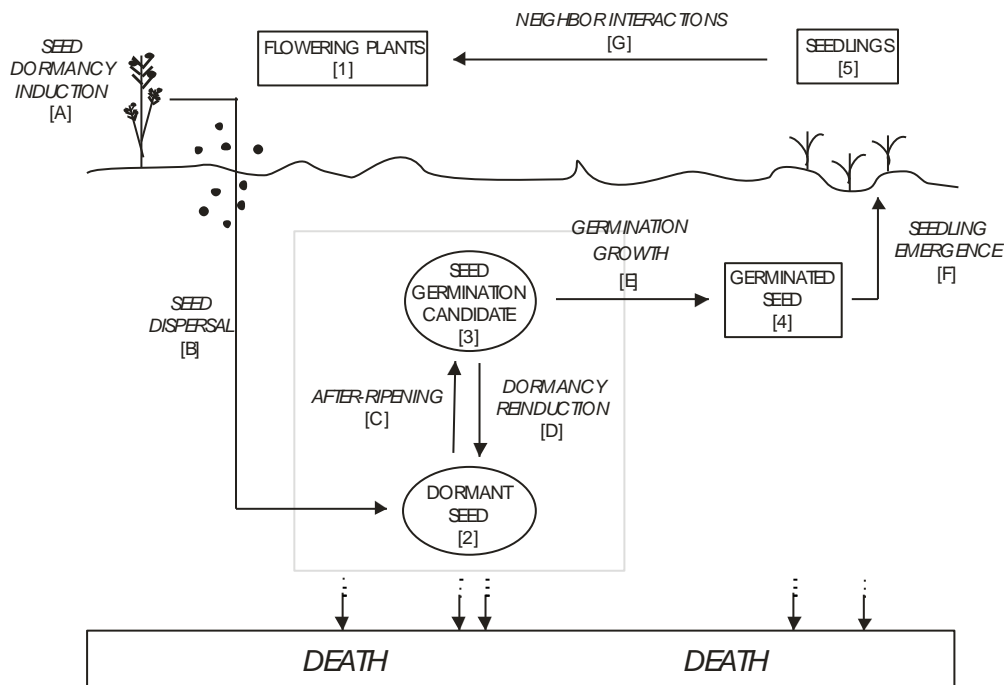

**Figure 6**. Schematic diagram of weedy *Setaria* sp. life history soil seed pool behavior: plant/seed state pools (1-5) and processes (A-F; C-D are reversible).

**Precondition for natural selection-elimination**. Excess local *Setaria* phenotypes compete for limited opportunity spacetime in a locality. Opportunity spacetime is the locally habitable space for an organism at a particular time which includes its resources (light, water, nutrients, gases) and conditions (heat, weather), disturbance history (e.g. tillage, herbicides, winter freezing), and neighboring organisms (e.g. crops, other weed species) (Dekker, 2009). The character of local spacetime seized and exploited by local populations of *Setaria* is typified by predictable disturbances in resource-rich cropping systems (e.g. Iowa, USA, maize-soybean fields; figure 12, top, bottom).

**Process of natural selection 1: produce phenotypic variation**. Variation in individual traits (hence individual fitness) is generated during seed fertilization and embryogenesis, and released at seed abscission. Local adaptation arises from natural selection and elimination among these variable phenotypes. Arguably the most crucial group of functional traits in generating phenotypic diversity induced during embryogenesis are those responsible for germinability-dormancy capacity heterogeneity (seed heteroblasty), the blueprint for seedling emergence timing (Jovaag, 2006). The key traits responsible for seed heteroblasty include differential development of three seed compartments enveloping the embryo (Dekker et al., 1996): seed hull shape (Dekker & Luschei, 2009; Donnelly et al., 2009), placental pore and seed transfer aleurone cell layer (TACL) membrane aperture qualities (Rost, 1971; Rost & Lersten, 1970), and the those of a putative oxygen-scavenging protein in the seed (Dekker and Hargrove, 2002). The light environment (photoperiod) of the flowering *Setaria* synflorence is the effective environmental signal modulating the development of these three morpho-physiological traits controlling seed heteroblasty (Atchison, 2001; Dekker, 2003). The traits affecting light interception include plant shoot-tiller architecture and individual seed position on the flowering synflorescence. Experimentally seed heteroblasty is determined by seed germination assays at abscission (figure 7; Atchison, 2001; Jovaag, 2006): after-ripening (AR; 4°C, moist, dark) followed by germination assay (e.g. 15-35°C, moist, light). Induction of heterogeneous seed dormancy occurs at several observable time scales: during the ca. 12d embryogenic period of individual seeds on a parent plant (figure 7, left; Dekker et al., 1996); and within and among populations with seasonal time (figure 7, right; table 2). The declining diurnal light period induces increasing germinability (decreasing dormancy) capacity in *S. faberi* seeds as time and photoperiod change (July to November).

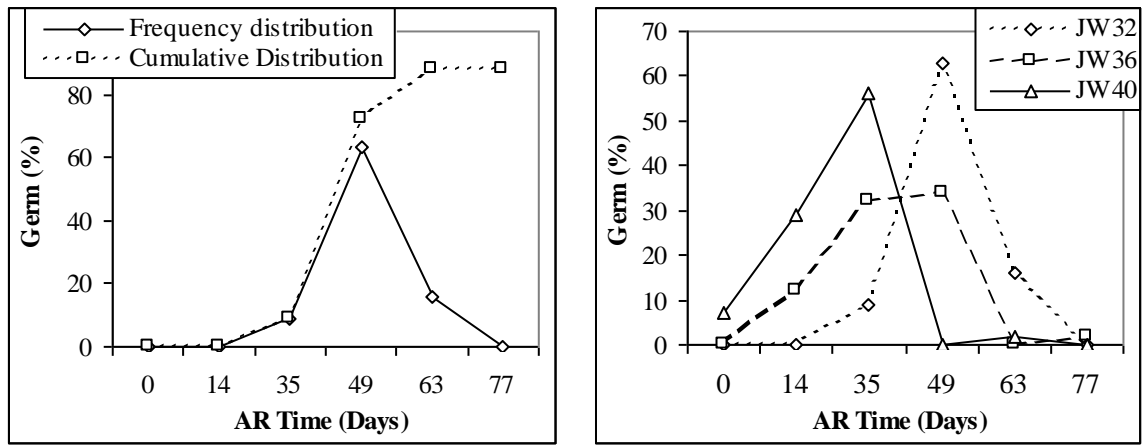

**Figure 7**. *S. faberii* seed germination heterogeneity among individual seeds of a single Ames, Iowa, USA population collected in Julian weeks (JW) 32, 36, and 40, 1998; left: JW 32, frequency/cumulative distributions; right: JW 32, 36, 38, frequency distribution.

| CONTRAST [1] | % GERM DIFFERENCE | |
|---|---|---|
| | 1998 | 1999 |
| Early - Middle | -37.8*** | -23.8*** |
| Early - Late | -58.6*** | -48.3*** |
| Middle - Late | -20.9*** | -24.4*** |

**Table 2**. Difference in *S. faberi* germination (% germ; least square mean) of four 1998 and 1999 populations collected during early (Julian week (JW) 32), middle (JW36) and late (JW40) seasonal periods. [1]ANOVA contrast, probability (p)>.05, ***=p<.001.

Phenotypic variation in a locality is also supplied by seed and pollen dispersal in space (gene flow) at metapopulation scales from landscape to global (population genetic structure; Wang et al., 1995a, b).

**Process of natural selection 2: survival and reproduction of the fittest phenotypes**. The second process of natural selection is the survival and reproduction of the variable phenotypes generated by the parent plant. FoxPatch represents this evolutionary process during the *Setaria* seed-seedling life history as the consequence environmental modulation of functional traits stimulating developmental change of seeds in the soil. Heteroblastic seeds begin their life with dispersal in both space and time. Dispersal in time is the formation of persistent soil seed pools in a locality. Seed states and processes of the local population in the soil are regulated by the interaction of three inherent morpho-physiological mechanisms with oxy-hydro-thermal-time (Dekker et al., 2003; Dekker and Hargrove, 2002). These trait-process interactions with soil signals are schematically presented in figure 8 (Dekker et al., 2003).

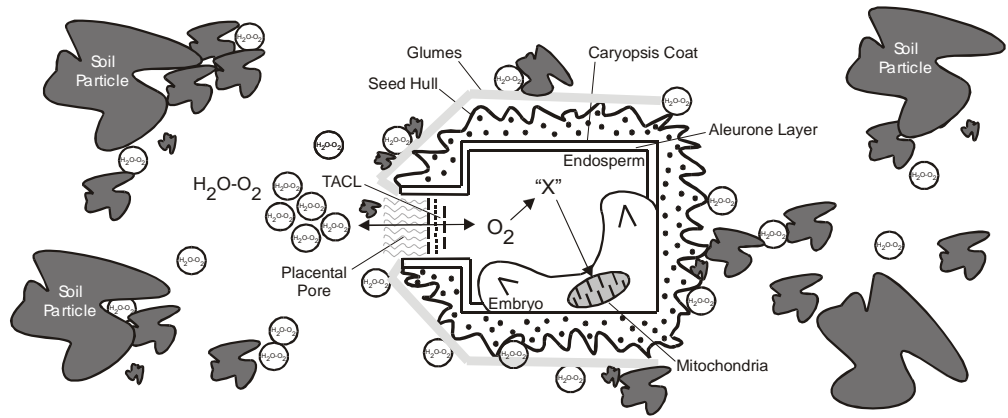

**Figure 8**. Schematic diagram of the *Setaria* sp. seed, surrounding soil particles and oxygen dissolved in water ($H_2O$-$O_2$). The seed interior (aleurone, TACL, endosperm, $O_2$-scavenging protein (X), embryo) is surrounded by the non-living glumes, hull, placental pore and the gas- and water-impermeable caryopsis coat.

The seed exterior hull acts as an environmental 'antenna' transducting soil signals (oxy-hydro-thermal-time) to the interior embryo. Soil-seed contact allows the accumulation and oxygenation of water on the rugose surface of weedy *Setaria* hull. Oxygenated water is channeled to the placental pore (hence into the interior embryo) by hull morphology. The role of seed hull morphology is apparent in the changes in shape and surface-to-volume ratios in weedy and domesticated *Setaria* species (figure 9; Donnelly et al., 2009). Soil contact, and the formation of oxygenated water films on the hull, play a crucial role in seed germination (Dekker and Luschei, 2009).

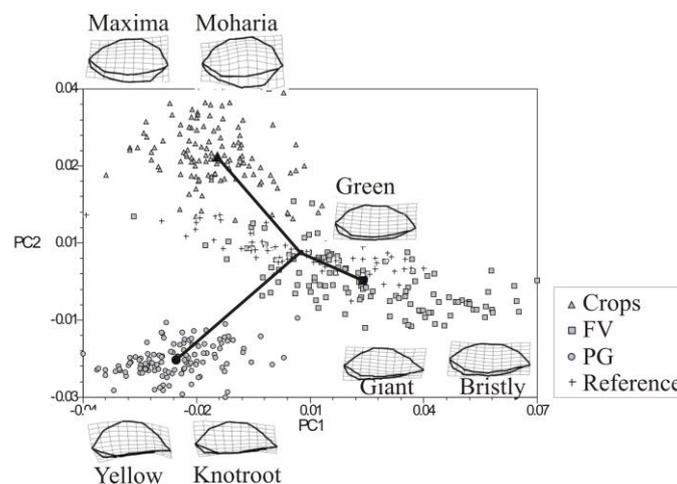

**Figure 9**. A principal components plot of the trajectory analysis for the lateral seed view. All vector magnitudes from the reference (*Setaria viridis* subsp. *viridis*) are significantly different. FV direction (*S. faberi*, *S. verticillata*) is significantly different from crops (*S. viridis* subsp. *italica* races maxima and moharia) and PG (*S. pumila*, *S. geniculata*). FV and PG are not significantly different.

The *Setaria* seed is surrounded by the caryopsis coat composed of several crushed cell layers (Rost, 1971). It is water- and gastight, and continuous except at the placental pore opening on the basal end of the seed. The mature seed is capable of freely imbibing water and dissolved gases, but entry is restricted and regulated by the placental pore and membrane control by the TACL (Rost & Lersten, 1970). Gases entering the moist seed interior must be dissolved in the imbibed water. This seed morphology strongly suggests that seed germination is restricted by water availability in the soil and by the amount of oxygen dissolved in water reaching the inside of the seed symplast to fuel metabolism (Dekker and Hargrove, 2002). Carbon monoxide (CO) stimulated germination in *S. faberi* has provided evidence of $O_2$-scavenging in the seed that delays or buffers the germination process: CO was found to poison this $O_2$ scavenging system (X) and thus speed the time until the critical, germination-threshold amount of $O_2$ is present in the symplast (Dekker and Hargrove, 2002; Sareini, 2002).

FoxPatch representation of soil environmental signal modulation of seed germinability-dormancy behavior. FoxPatch represents each *Setaria* seed process by a behavior rule, and an algorithmic prediction rule (Dekker et al., 2003). Rules for each life history process (C-E, figure 6) are a specification of these more general rules:

> **general seed behavior rule:** the behavior of an individual weedy *Setaria* seed in the soil is regulated by the amount of oxygen dissolved in water (the $O_2$-$H_2O$ signal) that accumulates in the seed symplast, and temperatures favorable (or not) to germination growth (the germination temperature signal), over some time period (cumulatively $O_2$-$H_2O$ -thermal time).

> **general prediction algorithm:** an individual weedy foxtail seed will change state when the minimum inherently-required $O_2$-$H_2O$-thermal-time signal is received from its realized environment (plus signals not causing an effect due to inefficient transduction or insensitivity).

The inherent germinability-dormancy capacity induced in an individual seed by the time of abscission (seed starting condition) can be experimentally determined in optimal conditions:

> **initial individual seed germinability-dormancy capacity**: the minimum $O_2$-$H_2O$-thermal-time signal required to stimulate germination at abscission

Each seed state change process can be experimentally determined in the field by frequent (e.g. hourly) measurement of soil temperature (thermal time; calculated $O_2$ solubility at temperature) and $H_2O$ content in the soil-seed profile.

Survive in the soil environment (dispersal in time). Seed in the soil cycle between two reversible states (dormant, germination candidate) until effective signals permit irreversible germination growth leading to seed germination. Seeds dispersed in the local soil seed pool remain alive until they emerge and begin autotrophic vegetative growth to exploit locally available opportunity spacetime. Typically *Setaria* seed germinability-dormancy cycles during the year: relatively greater $O_2$-$H_2O$-thermal-time signals increase germinability (e.g. cool moist spring), while relatively lesser signals reinduce secondary dormancy (e.g. hot dry summer) (Figure 10, Jovaag, 2006).

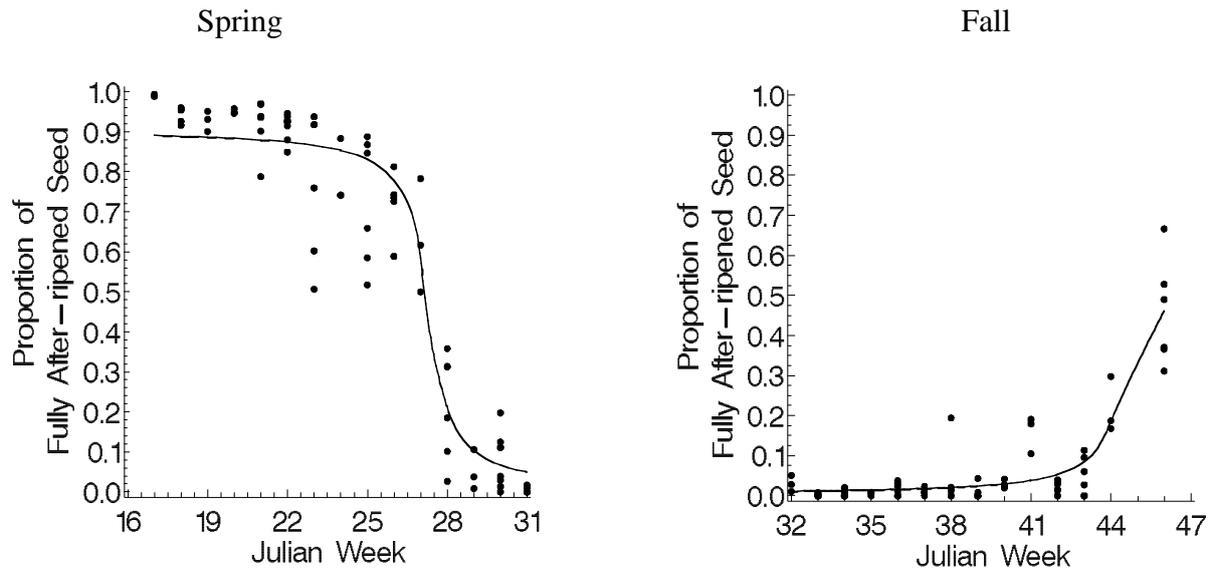

**Figure 10**. Proportion of highly germinable seed (fully after-ripened) versus Julian week (JW) for spring (JW 16-31, left) and fall (JW 32-47, right), first year after burial of four *S. faberi* Iowa, USA, populations. Dots: individual replicate observations. Solid lines: fitted model (3 parameter Lorentzian functions with a power of the mean variance model.

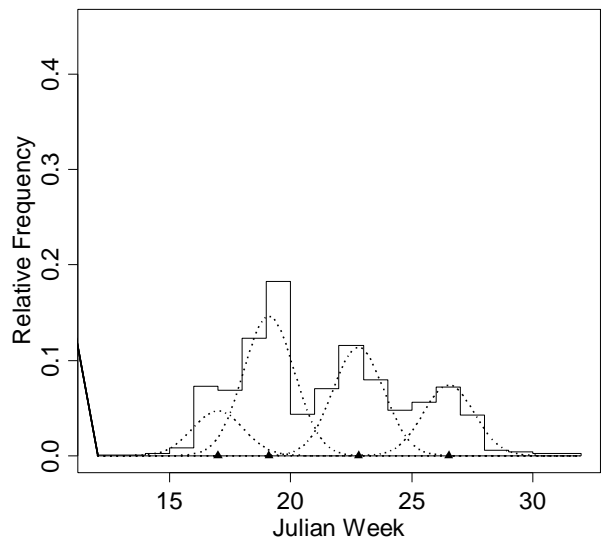

**Figure 11**. *S. faberi* seedling emergence (proportion of total) with time (Julian week, JW) during the spring and early summer of the first year after burial for all populations; bars: relative frequency; solid line: mixture model estimate (4 normal components with equal variance); dashed lines: model's 4 components weighted by the mixing proportions.

Emerge as a seedling at opportune seasonal times. Seedling recruitment timing is the single most important determinant of the subsequent interactions between an individual phenotype and its neighbors in a local community that directly determine survival and reproduction. Seed in the soil reversibly cycle between dormant and germination candidate states until conditions permit irreversible germination growth leading to germinated seed. As with all living seed processes in the soil, the effective signal stimulating germinative growth in heterogeneous seed is $O_2$-$H_2O$-thermal-time.  Germinated seed either emerge as seedlings or they experience fatal germination (mortality).

Complex oscillating patterns of *S. faberi* seedling emergence were observed during the first half of the growing season in all 503 soil burial cores of the 39 locally adapted *S. faberi* populations studied in central Iowa, USA, maize-soybean cropping systems (figure 11; Jovaag, 2006).  These characteristic patterns were attributed to six distinct dormancy phenotype cohorts arising from inherent somatic polymorphism in seed dormancy states.  The resulting pattern of emergence revealed the actual "hedge-bet" structure for *S. faberi* seedling recruitment investment, its realized niche, an adaptation to generous resource availability, mortality risks (especially those from predictable cropping system disturbances), and interactions with neighbours in those agroecosystems.

Fitness in *S. faberi* is conferred by strategic diversification of seedling recruitment. Evolutionarily, hedge-betting is a strategy of spreading risks to reduce the variance in fitness, even though this reduces intrinsic mean fitness.  These complex patterns in seedling recruitment behaviour support the conjecture that the inherent dormancy capacities of *S. faberi* seeds provides a germinability 'memory', preadaptation, of successful historical exploitation of local opportunity spacetime.  Seed heteroblasty is the inherent starting condition that interacts in both a deterministic and plastic manner with effective environmental signals to define the consequential heterogeneous life history trajectories of the population.

| Month | APRIL | | | | MAY | | | | | JUNE | | | | JULY | | | | | AUGUST | | | | SEPTEMBER | | | | OCTOBER | | | | NOVEMBER | | | | DEC |
|---|---|---|---|---|---|---|---|---|---|---|---|---|---|---|---|---|---|---|---|---|---|---|---|---|---|---|---|---|---|---|---|---|---|---|---|
| Julian Week | 15 | 16 | 17 | 18 | 19 | 20 | 21 | 22 | 23 | 24 | 25 | 26 | 27 | 28 | 29 | 30 | 31 | 32 | 33 | 34 | 35 | 36 | 37 | 38 | 39 | 40 | 41 | 42 | 43 | 44 | 45 | 46 | 47 | 48 | 49 | 50 |

Maize
- Seedbed Prep
- Planting
- Weed Control
- Layby
- Harvest
- Autumn Tillage

*S. faberi* Recruitment Cohorts
- Early Spring: 12
- Mid-Spring: 38
- Late Spring: 30
- Early Summer: 20
- Summer: 0.2
- Autumn: 0.1

Soybeans
- Seedbed Prep
- Planting
- Weed Control
- Layby
- Harvest
- Autumn Tillage

**Figure 12**. Calendar of historical, seasonal times (Julian week, month) of agricultural field disturbances (seedbed preparation; planting; weed control, including tillage and herbicides; time after which all cropping operations cease, layby; harvest and autumn tillage), and seedling emergence timing for central and southeastern Iowa, US, *Setaria faberi* population cohorts (all *S. faberi* combined: time, +/- S.E.; mean proportion; see table 3) (Jovaag thesis)

There exist predictable sources of mortality and recruitment opportunities for these *S. faberi* populations in Iowa, USA, agroecosystems over the course of their annual life histories (table III). The majority of seedlings were recruited in the spring when the risk of mortality is very high from crop establishment practices (seedbed preparation, planting, weed control) and the fecundity potential is the greatest. Weed seedlings emerging early have the greatest time available for biomass accumulation and competitive exclusion of later emerging neighbors. Subsequent fitness devolves on those individual *S. faberi* plants that escape these disturbances (Jovaag, 2006). As seasonal seedling recruitment proceeds potential fecundity and risk change. These factors result in differential seedling recruitment investment and strategy among the remaining emergence cohorts in response to changing opportunity spacetime (figure 12, table 3).

| COHORT | 1 | 2 | 3 | 4 | 5 | 6 |
|---|---|---|---|---|---|---|
| SEASON | EARLY | | | | LATE | |
| | Early Spring | Mid-Spring | Late Spring | Early Summer | Summer | Autumn |
| TIME (JW) | 16-18 | 18-20 | 22-24 | 26-28 | 32-35 | 45-49 |
| Fecundity Potential | very high | very high | high | medium | low | very low |
| Mortality Risk | very high | very high | high | low | low | low |
| Source(s) of Risk | crop disturbance | crop disturbance | crop disturbance | neighbors | neighbors | crop disturbance; climate |
| Weed Strategy | escape cropping | escape cropping | escape; post-layby opportunity | post-layby opportunity | post-layby opportunity | post-harvest opportunity |
| Seedling Investment | 12% | 38% | 30% | 20% | 0.2% | 0.1% |

**Table 3**. *S. faberi* seedling recruitment cohort (time, Julian week (JW)) exploitation of changing opportunity spacetime in Iowa, USA, maize-soybean cropping fields.

There exists a relationship between seed heteroblasty at abscission and its subsequent behavior in the soil that can be exploited to predict recruitment pattern: seed heteroblasty 'blueprints' seedling recruitment. Seedling numbers and temporal emergence patterns exploit local opportunity spacetime (Jovaag, 2006). Evidence of this relationship between heteroblasty and emergence numbers was provided by the positive Spearman correlation between dormancy capacity at abscission and the cumulative number of seeds emerged during the first year after burial for both the 1998 and 1999 *S. faberi* populations. Additionally, more dormant populations had lower emergence numbers during the first year after burial than less dormant populations. Early maturing seed was the most dormant and had the least number of seeds emerging. Seed maturing late in the season was the least dormant and had the greatest number of seeds emerging.